\documentclass{PoS}

\usepackage{url}

\title{Search for invisible decays of the Higgs boson at the ILC}

\ShortTitle{Search for invisible decays of the Higgs boson at the ILC}

\author{
  \speaker{Akimasa Ishikawa}\\
  High Energy Accelerator Research Organization (KEK), Tsukuba 305-0801,\\
  SOKENDAI (The Graduate University for Advanced Studies), Hayama 240-0193,\\ 
  E-mail: \email{akimasa.ishikawa@kek.jp}
}

\abstract{The existence of dark matter has been established in astrophysics. However, there are no dark matter candidates in the Standard Model~(SM). If the dark matter particles or their mediator can not interact with SM fermions or gauge bosons, the Higgs boson is the only portal to the dark matter. We present a simulation study to search for invisible decays of the Higgs boson at the ILC with the ILD detector.}

\FullConference{XXIX International Symposium on Lepton Photon Interactions at High Energies - LeptonPhoton2019\\
		August 5-10, 2019\\
		Toronto, Canada}

\begin{document}

\section{Introduction}
The existence of dark matter has been established in astrophysics; the rotation curve of a disk galaxy, spacial distributions of luminous baryonic matter and total matter in a collision of galaxy clusters, and an anisotropy spectrum of cosmic microwave background~\cite{Aghanim:2018eyx}, baryon acoustic oscillations~\cite{Delubac:2014aqe}, and large scale structure of the Universe.  
However, there are no dark matter candidates in the Standard Model~(SM). 
If the dark matter particles or their mediator can not interact with SM fermions or gauge bosons, the Higgs boson is the only portal to the dark matter. We present a simulation study to search for invisible decays of the Higgs boson at the International Linear Collider~(ILC) with the International Large Detector~(ILD) detector.

\section{ILC and ILD}
The ILC is an electron-positron collider. The center-of-mass energy~($E_{CM}$) is 250~GeV at the first stage and upgradable to at least 1~TeV at Kitakami site. The length of the ILC is about 20km for the first stage and 50km for the 1~TeV upgrade. The ILC has a beam polarization capability of $\pm80$\% and $\pm30$\% for electron and positron beams, respectively. This is a powerful tool to select the quantum number of intermediate state and to suppress backgrounds. 
The ILD is one of the two detectors at the ILC experiment. The ILD detector is optimized to particle flow algorithm~\cite{Brient:2002gh}. All subdetectors, especially calorimeter system, are highly segmented to enable particle flow approach to maximize the jet energy resolution.

\section{Run Scenario}
The run scenario assumed in this manuscript is based on H-20 scenario~\cite{Barklow:2015tja}. The $E_{CM}$ is initially 250~GeV and upgraded to 350~GeV and 500~GeV. At all three $E_{CM}$, two polarization configuration are used $P(e^-, e^+)=(-80\%, +30\%)$ and $P(e^-, e^+)=(+80\%, -30\%)$, named ``Left'' and ``Right'', respectively in this manuscript. Table~\ref{tab:lumi} summarizes the integrated luminosities.

\begin{table}[hb]
\begin{center}
\caption{Integrated Luminosities with H-20 scenario.}
\label{tab:lumi}
\begin{tabular}{l|rr}
\hline
$E_{CM}$ [GeV]       & ``Left'' [$fb^{-1}$] & ``Right'' [$fb^{-1}$] \\
\hline
250     & 1350 &  450 \\
350     &  135 &   45 \\
500     & 1600 & 1600 \\
\hline
\end{tabular}
\end{center}
\end{table}

\section{Reconstruction}
\begin{figure}[htb]
  \begin{center}
    \includegraphics[width=0.45\textwidth]{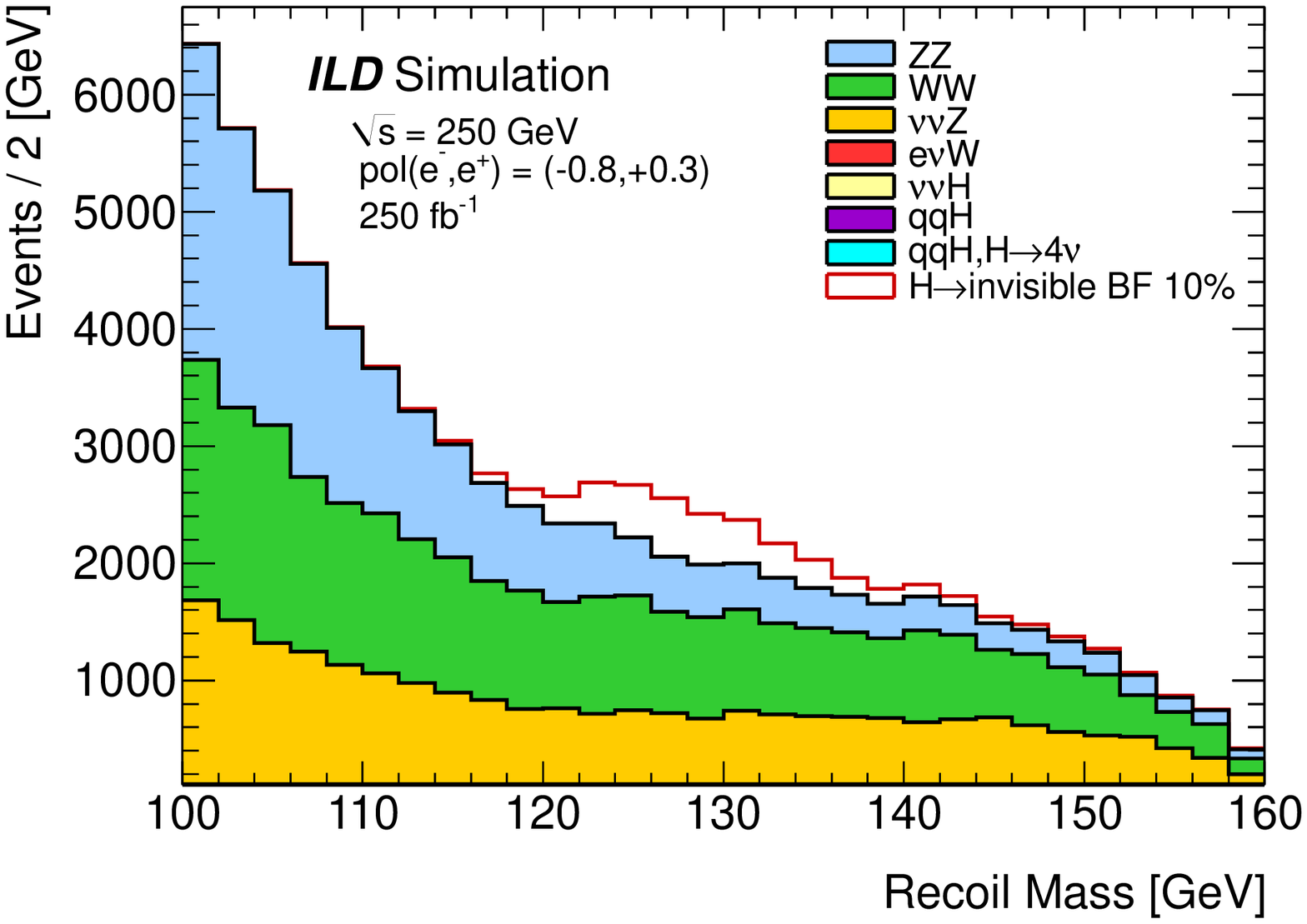}
    \includegraphics[width=0.45\textwidth]{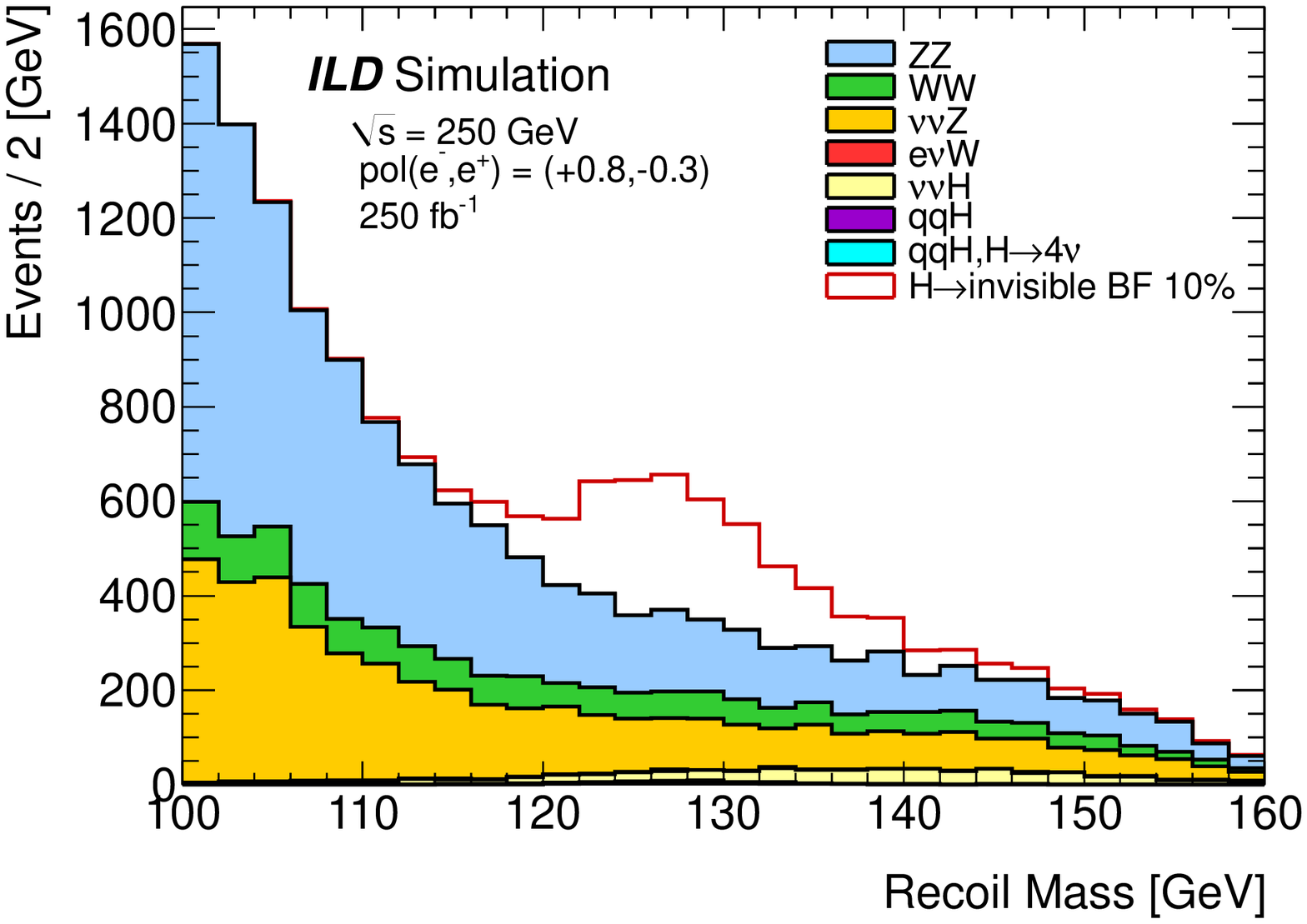}
    \caption{Recoil mass distributions of $Z \to q\bar{q}$ for ``Left'' and ``Right'' polarizations.}
    \label{fig:recoilmass}
  \end{center}
\end{figure}

Invisible decays of the Higgs boson can be searched for using a recoil mass technique in $e^+e^- \to ZH$ processes in a model independent way. The initial state $e^+e^-$ is known since the electron and positron are both elementary and their four momenta are precisely determined by the beam-delivery system. The $Z$ boson is reconstructed from either of $q\bar{q}$ or $\ell^+\ell^-$ ($\ell$ stands for electron or muon) final states. 
The Higgs mass can be reconstructed as recoil mass against $Z$, $m_{\rm rec}^2 = P_{H}^2 = (P_{e^+e^-} - P_{Z})^2$. Dominant backgrounds are $e^+e^- \to ZZ, WW$ and $Z\nu\bar{\nu}$ for $Z \to q\bar{q}$. Figure~\ref{fig:recoilmass} shows the recoil mass distributions of $Z \to q\bar{q}$ decays for ``Left'' and ``Right'' polarizations~\cite{ishikawa}. It is notable that the ``Right'' polarization is suppressing numerous charge-current background processes due to chiral nature of the charge-current interaction.

\section{Upper Limits on the Branching Fraction}
From the recoil mass distributions for signal and backgrounds, we performed ensemble tests to estimate the upper limit on branching fraction of invisible decays of the Higgs boson. The SM process $H \to ZZ^* \to 4\nu$ is also invisible final state and gives the same recoil mass distribution. This known contribution is separately considered in the fit in ensemble test and is not included to the upper limit for the invisible decays of Higgs.
The $Z \to q\bar{q}$~\cite{ishikawa} and $Z \to \ell^+\ell^-$~\cite{tian} are separately analyzed and then combined.  Table~\ref{tab:resultstime} shows results of upper limits with $Z \to q\bar{q}$ and combined $q\bar{q}$ and $\ell^+\ell^-$ cases for each $E_{CM}$ with the same running time. The integrated luminosities of 250~$fb^{-1}$, 350~$fb^{-1}$ and 500~$fb^{-1}$ are taken at $E_{CM}$ of 250~GeV, 350~GeV and 500~GeV, respectively.
The upper limit with $q\bar{q}$ final states is much better than that with $\ell^+\ell^-$ thanks to the larger branching fraction of hadronic $Z$ decays.
Combination with the result from pure leptonic final state improves precision at all center of mass energies.
Lower $E_{CM}$ gives better results since the cross section of $e^+e^- \to ZH$ is larger at smaller center of mass energies.. The upper limit with ``Right'' polarization is better than that with ``Left'' polarization since dominant backgrounds are suppressed. In summary, operation at $E_{CM}$ of 250~GeV with ``Right'' polarization gives the best upper limit with the same running time.

\begin{table}[htb]
\begin{center}
\caption{Upper limits on the branching fraction of invisible decays of the Higgs boson in percent assuming the same running time for each $E_{CM}$ corresponding to 250~$fb^{-1}$, 350~$fb^{-1}$ and 500~$fb^{-1}$ at 250~GeV, 350~GeV and 500~GeV, respectively.}
\label{tab:resultstime}
\begin{tabular}{l|rr|rr}
\hline
$E_{CM}$ [GeV]   & \multicolumn{2}{c|}{$Z \to q\bar{q}$}    & \multicolumn{2}{c}{$Z \to q\bar{q}$ and $\ell^+ \ell^-$} \\ 
                & ``Left''  & ``Right''   & ``Left''  & ``Right''     \\
\hline
250    & 0.95 &  0.65 & 0.86 & 0.61 \\
350    & 1.49 &  1.37 & 1.23 & 1.10 \\
500    & 3.16 &  2.30 & 2.39 & 1.73 \\
\hline
\end{tabular}
\end{center}
\end{table}

Table~\ref{tab:results} shows the upper limits based on H-20 scenario. By combining all runs, an upper limit of 0.26\% can be achieved at the ILC. This result can be compared with the current model dependent LHC results of 26\% and 19\% by ATLAS~\cite{atlas} and CMS~\cite{cms}, respectively, and 3.8\% at HL-LHC with 3~$ab^{-1}$~\cite{Cepeda:2019klc}. The ILC will give two orders~(one order) of magnitude better result than current LHC (future HL-LHC) experiments.

\begin{table}[htb]
\begin{center}
\caption{Upper limits on the branching fraction of invisible decays of the Higgs boson in percent based on H-20 scenario.}
\label{tab:results}
\begin{tabular}{l|rrr|rrr}
\hline
$E_{CM}$ [GeV]   & \multicolumn{3}{c|}{$Z \to q\bar{q}$}            & \multicolumn{3}{c}{$Z \to q\bar{q}$ and $\ell^+ \ell^-$}\\ 
               & ``Left''  & ``Right''   & combined & ``Left''  & ``Right''    & combined\\
\hline
250      & 0.41 &  0.51 & 0.32 & 0.37 & 0.45 & 0.28\\
350      & 2.40 &  3.82 & 2.03 & 1.98 & 3.07 & 1.66\\
500      & 1.77 &  1.29 & 1.04 & 1.34 & 0.97 & 0.79\\
\hline
combined & --   & --    & 0.30 & --   & --   & 0.26\\
\hline
\end{tabular}
\end{center}
\end{table}

\section{Conclusion}
The ILC is the ideal place to search for invisible decays of the Higgs boson with model independent recoil mass technique. With H-20 scenario, an upper limit on branching fraction of Higgs decaying to invisible final state of 0.26\% can be achieved.

\section{Acknowledgment}
A.~I. is supported in part by the Japan Society for the Promotion of Science Grant No.~16H02176.

\end{document}